\documentclass[aps,prb,twocolumn,showpacs,superscriptaddress]{revtex4}
\usepackage{CJK}
\usepackage{bm}
\usepackage{graphicx}
\usepackage{color}
\usepackage{amsmath,amssymb,amsfonts,amsthm}

\def \beq {\begin{equation}}
\def \eeq {\end{equation}}

\def \bmatrix {\begin{pmatrix}}
\def \ematrix {\end{pmatrix}}

\def \H {\mathcal{H}}
\def \E {\mathcal{E}}

\def \T {\hat{T}}

\def \S {\hat{S}}
\def \P {\hat{P}}
\def \K {\hat{\mathcal{K}}}
\def \i {\hat{i}}

\def \Z {\mathbb{Z}}

\begin{document}

\title{Nonsymmorphic symmetry-required band crossings in topological semimetals}

\author{Y. X. Zhao}
\email[]{y.zhao@fkf.mpg.de}
\affiliation{Max-Planck-Institute for Solid State Research, D-70569 Stuttgart, Germany}

\author{Andreas P. Schnyder}
\email[]{a.schnyder@fkf.mpg.de}
\affiliation{Max-Planck-Institute for Solid State Research, D-70569 Stuttgart, Germany}

\begin{abstract}
We show that for two-band 
systems  nonsymmorphic symmetries may enforce the existence of band crossings in the bulk, which 
realize Fermi surfaces of reduced dimensionality.
We find that these unavoidable crossings originate from the momentum dependence of the nonsymmorphic symmetry, which puts strong restrictions on the global structure of the band configurations. 
Three different types of nonsymmorphic symmetries are considered: (i) a unitary nonsymmorphic symmetry, (ii) a nonsymmorphic magnetic symmetry,
and (iii) a nonsymmorphic symmetry combined with inversion.
For nonsymmorphic symmetries of the latter two types, the band crossings are located at high-symmetry points of the Brillouin zone, with their exact positions being determined by the algebra of the symmetry operators.
To characterize these band degeneracies we introduce a \emph{global} topological charge and show that it is
of $\mathbb{Z}_2$ type, which is in contrast to the \emph{local} topological charge of Fermi points in,
say, Weyl semimetals.
To illustrate these concepts, we discuss the $\pi$-flux state  as well as the  
SSH model at its critical point and show that these two models
fit nicely into our general framework of nonsymmorphic two-band systems.


\end{abstract}

\date{\today}



\maketitle

\section{Introduction}
Since the experimental discovery of topological insulators~\cite{MarkusKonig11022007,Hsieh:2008fk}, symmetry protected topological  
phases have become a major research subject~\cite{hasan:rmp,qi:rmp_backup,hasan_moore_review,schnyderReviewTopNodalSCs,Senthil2014,chiu_review_arxiv}.
Recent studies have been concerned with topological  phases that are protected by spatial symmetries,
such as topological crystalline insulators~\cite{Ando_Fu_TCI_review_backup,Chiu_reflection,Sato_Crystalline_PRB14,Morimoto2013}
and topological semimetals stabilized by reflection, inversion, or other crystal symmetries~\cite{ChiuSchnyder14,Nodal_line_chan,zhao_PRL_16}. 
Until recently, the study of these topological crystalline materials has focused on the role of point group symmetries. However, besides point
group symmetries the space group of a crystal can also contain nonsymmorphic symmetries, which are combinations of point group operations with
nonprimitive lattice translations. It has been shown that the presence of nonsymmorphic symmetries leads
to new topological phases, which can be  insulating~\cite{nonsymmorphic_Sato,New_crystalline_Fang,nonsymmorphic_Liu,2D_classification_Liu,photonic_TI_Fu,shiozaki_gomi_arXiv_16},
or semimetallic with Dirac points protected by  nonsymmorphic symmetries~\cite{nonsymmorphic_cone_Kane,tomas_arXiv_1604}.
In the latter case, the Dirac points possess  \emph{local} topological charges, which guarantees their local stability.

However, as we show in this paper, nonsymmorphic symmetries restrict the form of the band structure not only
locally but also \emph{globally}, which may lead to unavoidable band crossings in the bulk~\cite{Michel2001377,konig_mermin_PRB_97,watanabe_vishwanath_arXiv_16,parameswaran_arXiv_15,parameswaran_nat_phys_13}. 
Indeed, the nonsymmorphic symmetries can put so strong constraints on the global properties of the band structure that 
 the system is required by symmetry to
be in a topological semimetal phase, with Fermi surfaces of reduced dimensionality.
These symmetry-enforced semimetals possess low-energy excitations with unconventional dispersions
and may exhibit novel topological response phenomena
and unusual magneto-transport properties. 
In the following we consider three different types of nonsymmorphic symmetries: (i)
Unitary nonsymmorphic symmetries, (ii) nonsymmorphic symmetries combined with inversion, and
(iii) nonsymmorphic magnetic symmetries.
We first rigorously  prove that for any one-dimensional (1D) two-band system  unitary nonsymmorphic symmetries enforce the existence of band crossings, due to global topological constraints on the band structure. 
In the presence of an additional inversion symmetry, the symmetry enforced band degeneracies  
are located either at the origin or at the boundary of the Brillouin zone (BZ), depending on the algebra of the symmetry operators.  
The same holds true for nonsymmorphic magnetic symmetries, which are composed of a
unitary nonsymmorphic symmetry followed by an anti-unitary time-reversal symmetry. 
We present generalizations of these results to higher dimensions, for which nonsymmorphic symmetries may enforce the existence 
of zero- or higher-dimensional band crossings.
In all of the above cases we find that 
the nonsymmorphic symmetries restrict the momentum space structure
in the  BZ both locally and globally.
To characterize the global topological features we introduce a novel global topological charge, which
as we show, is always of $\mathbb{Z}_2$ type.
Hence, the global topological features exhibit a $\mathbb{Z}_2$ classification,
which is in contrast to the local topological characteristics, which
possess a  $\mathbb{Z}$ classification. 
Finally, we illustrate these findings by considering two prototypical examples: 
(i) the $\pi$-flux square lattice model and (ii) the
 SSH model at its critical point. 
Within our unified framework, we show that the former model can
be viewed as the higher-dimensional generalization of the latter.



%

\begin{figure}
	\includegraphics[scale=0.5]{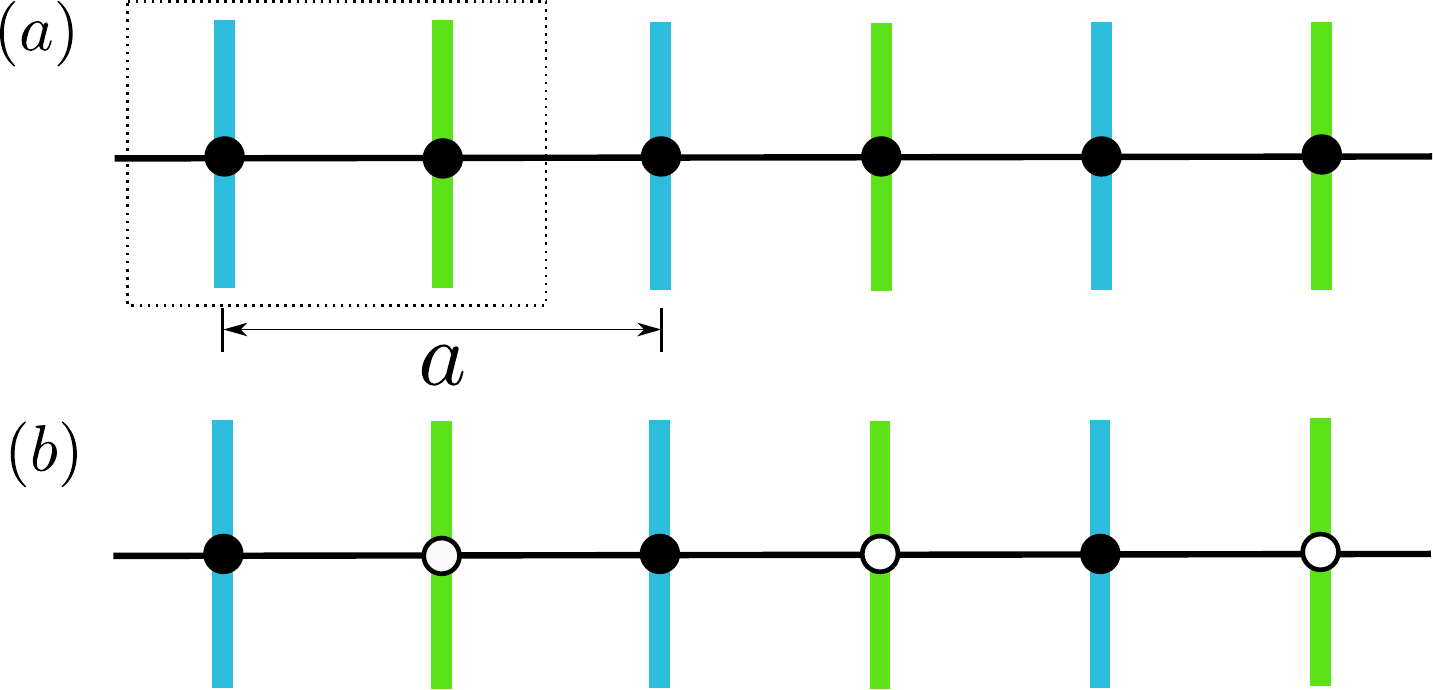}
	\caption{ \label{mFig1}
	Illustration of nonsymmorphic symmetries in one-dimensional lattices. (a) The nonsymmorphic symmetry is composed of 
	a  $\pi$ rotation 
	followed by a half translation $a$/2, where $a$ is the lattice constant. (b) The nonsymmorphic \emph{magnetic} symmetry is composed of
	the two operations of (a) followed by the exchange of black and white balls, which represents time-reversal symmetry.  \label{Non-sym}}
\end{figure}

\section{Unitary nonsymmorphic symmetry}
We start by considering a general 1D two-band Hamiltonian $\H(k)$ with the two-fold unitary nonsymmorphic symmetry 
\beq  G(k)=\bmatrix 0 & e^{-ik}\\ 1 & 0 \ematrix ,  \label{Nonsymm-operator} \eeq
which acts on $\H(k)$ as
\beq G(k) \H(k) G^{-1}(k)=\H(k) . \label{Non-symm} \eeq
Since 
$G^2(k)= e^{-ik}\sigma_0$, 
the eigenvalues of $G(k)$ are $\pm e^{-i k/2}$.
Therefore, the nonsymmorphic symmetry $G(k)$ can be viewed as an operation on internal degrees of freedom (e.g., pseudospin) followed by a half translation, as illustrated in Fig.~\ref{mFig1}(a).
Observing that $G(k)$ anti-commutes with $\sigma_3$, the Hamiltonian can be written as
\beq
\H(k)=\bmatrix 0 & q(k)\\ q^*(k) & 0\ematrix. \label{Chiral-form}
\eeq
Without loss of generality, we have dropped the term proportional to the identity, which only shifts the energy of eigenstates.
Inserting Eqs.~(\ref{Chiral-form}) and (\ref{Nonsymm-operator}) into Eq.~(\ref{Non-sym}), we find
that due to the non-symmorphic symmetry $G(k)$,
$q(k)$ must satisfy
\beq q(k)e^{ik}=q^*(k) . \label{Non-sym-q}\eeq
We claim that any periodic function $q(k)$ satisfying Eq.~(\ref{Non-sym-q}) has zeros, and thus any two-band model with the nonsymmorphic symmetry (\ref{Nonsymm-operator}) is required to be gapless.
To see this, we introduce $f(z)=q(k)$ with $z=e^{ik}$, from which it follows that $zf(z)=f^*(z)$. If $q(k)$ or $f(z)$ is nonzero on the unit circle $S^1$, then
\begin{eqnarray} \label{zFz}
z=f^*(z)/f(z) ,
\end{eqnarray} 
 which however is impossible. This is because for $z\in S^1$, the two sides of Eq.~\eqref{zFz} both define functions from $S^1$ to $S^1$,
but the left-hand side has winding number $1$, while the winding of the right-hand side is even, since $f^*(z)/f(z)=e^{2i\mathrm{Arc}[f(z)]}$. Thus, $q(k)$ must vanish at some momentum by contradiction. For the topological argument to work for multi-band theories, we may replace $q(k)$ in Eq.(\ref{Chiral-form}) by the determinant of the off-diagonal entry, which is discussed in Sec.\ref{Discussion}.

\begin{figure*}
	\includegraphics[scale=1.2]{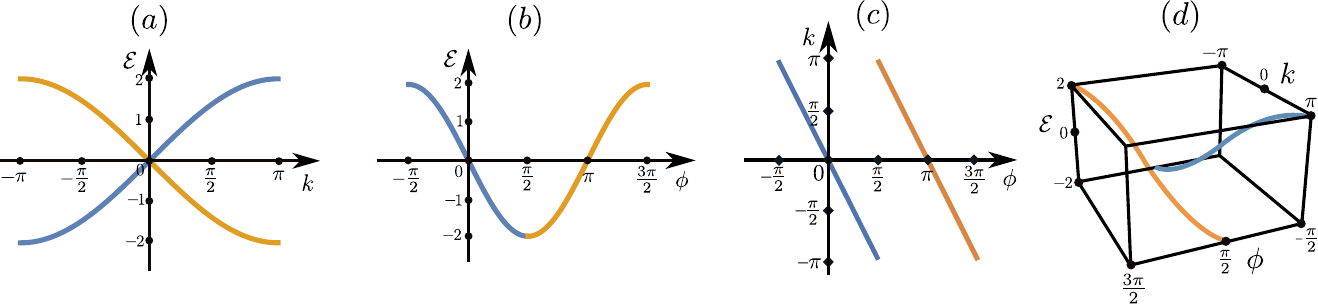}
	\caption{(a) Energy spectrum  $\E_\pm(k)$ of Hamiltonian~\eqref{Simplest-model}. Blue and orange
		correspond to the eigenstates $\E_+$ and $\E_-$, respectively. The two eigenstates are connected smoothly at the boundary of the BZ
		and cross each other at the center of the BZ.
		(b) $\E_{\pm} (k) $ as a function of the phases $\phi$ of the nonsymmorphic symmetry eigenvalue $g_{\pm}(k)$. In the space of the eigenvalues $g_{\pm} (k)$ of the nonsymmorphic symmetry, the two bands are smoothly connected with each other, without any crossing point. (c) $k$ as a function of the phase $\phi$ of the eigenvalues $g_{\pm} (k)$.  The two eigenvalue branches are connected at $\phi=\frac{\pi}{2}$ and $\phi=\frac{3\pi}{2}$ (= $-\frac{\pi}{2}$), leading to a winding number 2. (d)  Trajectory of the two bands in the $(k,\phi,\E)$ space. As a problem of essential three parameters, the two bands are connected as a circle in the $(k,\phi,\E)$ space, corresponding to $(2,1)\in H_1(S^1\times S^1\times \mathbb{R})\cong \mathbb{Z}\oplus \mathbb{Z}$.
		\label{spectra}
	}
\end{figure*}

\section{Nonsymmorphic symmetry combined with inversion symmetry}
We note that while a unitary nonsymmorphic symmetry guarantees the existence of a band crossing point,
it does not fix the position of this degeneracy point in momentum space.
However, in the presence of an additional inversion symmetry, the band crossings are pinned to either the origin or the boundary of the BZ.
To demonstrate this, let us consider the
 inversion symmetry $\P=\sigma_2\i$, where $\i$ inverses the momentum.
 We find that
\beq   
[\H,\P]=0, \quad \P G(k)\P^{-1}=-G^{T}(-k), \quad \P^2=-1. \label{relation_first_P} 
\eeq
Since $q(k)$ is a periodic function, we expand it as
$  q(k)=\sum_n q_n e^{ink}  $.
It follows from Eq.~(\ref{Non-sym-q}) that
$ q_{-(n+1)}=q^*_n, $
which, as a recursion relation, allows us to express $q(k)$ as
\begin{eqnarray} \label{q_expand_exp}
q(k)=\sum_{n=0}^\infty (q_n e^{ink}+q_n^* e^{-i(n+1)k}).
\end{eqnarray} 
From Eq.~\eqref{relation_first_P} it follows
that  $ \sigma_2 \H(-k)\sigma_2=\H(k), $
which implies $q(k)=-q^*(-k)$ or equivalently $ q_n=-q^*_{n}$.
Since $q_n$ are all purely imaginary, we find that
\beq q(k)=\sum_{n=0}^\infty \frac{\lambda_n}{i} ( e^{ink}- e^{-i(n+1)k}), \label{Seq-2} \eeq 
with $\lambda_n$ being real numbers.
We observe that independent of $\lambda_n$ there   always exists a band crossing point 
at $k=0$. 
For example, by keeping only the zeroth term in Eq.~(\ref{Seq-2}), one finds as a simple concrete model,
\beq \H_0(k)=\lambda \sin k \sigma_x+\lambda(1-\cos k)\sigma_y . \label{Simplest-model}\eeq
We note that the nonsymmorphic symmetry $G(k)$ relates seemingly independent terms to each other in the Hamiltonian.
This is exemplified by  Eq.~(\ref{Simplest-model}), where all three terms (which are usually independent) have the same coefficients. 
Obviously, higher order terms in Eq.~(\ref{Seq-2}), which constitute symmetry-preserving perturbations, cannot split the band crossing point 
of $\H_0$ at $k=0$. That is, the gapless mode at $k=0$ described by
the low-energy effective Hamiltonian $ \H_{\textrm{eff}}(k)=\lambda k \sigma_x $
is stable against symmetry-preserving perturbations. 

The fact that the Hamiltonian given by Eq.\eqref{Seq-2} exhibits a band crossing at $k=0$ can directly be seen by
computing the eigenstate of $G(k)$ and $\H(k)$.
Because $G(k)$ and $\H(k)$ commute [see Eq.\eqref{Non-symm}],  they can be simultaneously diagonalized by the same set of eigenstates
\beq \H(k)|\pm,k\rangle=\E_{\pm}(k)|\pm,k\rangle,\quad G(k)|\pm,k\rangle=g_{\pm} (k)|\pm,k\rangle , \eeq
where the eigenfunctions  $ |\pm,k\rangle$ are given by
\beq |+,k\rangle=\frac{1}{\sqrt{2}}\bmatrix 1\\ e^{i\frac{k}{2}} \ematrix,\quad |-,k\rangle=\frac{1}{\sqrt{2}}\bmatrix 1\\ -e^{i\frac{k}{2}} \ematrix  \eeq
and the eigenvalues are
\beq \label{energy_spectrum}
\E_\pm(k)= \pm2 \sum_{n=0}^\infty  \lambda_n \sin \left(n k +\frac{k}{2} \right), \quad g_{\pm}(k)=\pm e^{-i\frac{k}{2}}.
\eeq
We find that the energy and nonsymmorphic symmetry eigenvalues of $|+,k\rangle$ 
at $k=\pm \pi$ are continuously connected to the corresponding eigenvalues of $|-,k\rangle$ at $k=\mp \pi$, see Fig.\ref{spectra}.
That is, we have
\beq \begin{cases}
	\E_{+}(-\pi)=\E_{-}(\pi)\\
	\E_{-}(-\pi)=\E_{+}(\pi)
\end{cases},\quad  \begin{cases}
g_{+}(-\pi)=g_{-}(\pi)\\
g_{-}(-\pi)=g_{+}(\pi)
\end{cases} .\eeq
We note that the eigenfunctions $|\pm,k\rangle$ become degenerate  in energy at  $k=0$ [i.e., $\E_+ (0) = \E_- (0)$], while their
nonsymmorpic symmetry eigenvalue remains non-degenerate at $k=0$ [i.e., $g_+ (0) \ne g_-(0)$].
Therefore the two bands $|\pm,k\rangle$ must cross each other.

To see the topological features of the band structure, we first note that the eigenvalues $g_\pm (k)$ of the nonsymmorphic symmetries $G$ form
a manifold as a function of momentum $k$.
That is, the eigenvalues $g_{\pm} (k)$  are multivalued functions of $k$, with different branches being smoothly connected.  Inversely, $k$ is a single-valued continuous function of the eigenvalues of the symmetry $G$.  For the two-fold nonsymmorphic symmetry~(1), the momentum $k\in S^1$ has winding number $2$ as a function of the eigenvalue $g_\pm (k)\in U(1)$, which indicates a nontrivial topology
[see Fig.\ref{spectra}(c)]. To better understand this nontrivial topology, it is instructive to draw the mutual dependence of the energy eigenvalues $\E_{\pm}$, the nonsymmorphic eigenvalues $g_\pm $, and the momentum $k$ in terms of a trajectory in the three-dimensional space $(k,\phi,\E)$.
For the two-band model~(9) this is shown in Fig.~2(d). The projections of this trajectory onto  the three orthogonal planes  
($\E$, $k$), ($\E$, $\phi$), and ($k$, $\phi$) are shown in Figs.\ref{spectra}(a), \ref{spectra}(b) and \ref{spectra}(c), respectively. We can see that the two bands $\E_{\pm}$ are connected as a circle in $(k,\phi,\E)$ space, corresponding to the element $(2,1)$ in the homology group $H_1(S^1\times S^1\times \mathbb{R},\mathbb{Z})\cong \mathbb{Z}\oplus\mathbb{Z}$.

Instead of $\P=\sigma_2\i$, another possible choice for 
 $\P$ is $\P=\sigma_1\i$ with the symmetry relations
\beq [\H,\P]=0, \quad \P G(k)\P^{-1}=G^{T}(-k), \quad \P^2=1.\eeq
With this choice, we find the  following relations for $q(k)$ and $q_n$, 
\begin{eqnarray}
q(k)=q^*(-k), \quad q_n=q_n^*  .
\end{eqnarray}
Using Eq.~\eqref{q_expand_exp}, it follows that
\beq q(k)=\sum_{n=0}^{\infty} \lambda_n(e^{ink}+e^{-i(n+1)k}). \label{Seq-3}\eeq
Hence, there always exists a band crossing point at $k=\pi$.

Let us now show that  the algebra obeyed by the symmetry operators determines whether
the band crossing point is at $k=0$ or $k= \pi$.
To that end, we recall that for the choice $\P=\sigma_2\i$   the operators at the inversion invariant point  $k=\pi$,  $\P=\sigma_2\i$, $G(\pi)=-i\sigma_2$, and $\H(\pi)$ are mutually commuting, see Eq.~\eqref{relation_first_P}.
At the other inversion invariant point $k=0$, however, $\P=\sigma_2\i$ and $G(0)=\sigma_1$ are anti-commuting, while $\H (0)$ commutes
with $\P$ and $G(0)$, i.e., $[\H(0),\P]=0$ and $[\H(0),G(0)]=0$.
It follows that the two degenerate eigenstates of  $\H$ at $k=0$ can be
written as eigenstates of $P$ with different eigenvalues.
Explicitly, we find that $\frac{1+i}{2}|+,0\rangle+\frac{1-i}{2}|-,0\rangle$ is an eigenstate of $\P$ 
with eigenvalue $+1$, while $\frac{1-i}{2}|+,0\rangle+\frac{1+i}{2}|-,0\rangle$ is an eigenstate of $\P$ with
eigenvalue $-1$. Therefore, the band crossing, which is protected by $\P$, occurs at $k=0$.

A similar analysis can be  performed for the choice  $\P=\sigma_1\i$, i.e., 
the Hamiltonian given by Eq.~(\ref{Seq-3}).
In that case, we find that at $k=0$ the operators $\H(0)$, $G(0)$, and $\P$ are mutually commuting,
while  $\P$ and $G(k)$ anti-commute at $k=\pi$, where the band degeneracy is located.
We conclude that the algebraic relations obeyed by the symmetry operators determine the location
of the symmetry-enforced band crossing, see Table~\ref{Positions}.


%


\section{Nonsymmorphic magnetic symmetry \label{MagneticNonsymmorphic}}
From the discussion in the previous section it follows that not all the symmetry constraints are necessary
to enforce the existence of the band crossing.
As we shall see, a single nonsymmorphic antiunitary symmetry, namely a magnetic nonsymmorphic symmetry, is sufficient to ensure 
the existence of a band crossing at $k=0$ or $k=\pi$.
As illustrated in Fig.~\ref{mFig1}(b), a magnetic nonsymmorphic symmetry can be viewed as the combination of a nonsymmorphic symmetry $G(k)$ 
with a time-reversal symmetry $\T$.
We only require that the combined symmetry $G \T$ is satisfied. That is, both $G$ and $\T$ may be broken individually,
but the combination must be preserved.
In what follows we assume that  $\T^2=+1$ and consider two possible choices for  $\T$, namely,
(i) $\T=\K\i$ and (ii) $\T=\sigma_3\K\i$, where $\K$ denotes the complex conjugation operator.
 By use of Eq.~\eqref{Nonsymm-operator}, we find that in case
(i) $[\T,G(k)]=0$, while in case (ii)  $\{\T,G(k)\}=0$.


Let us start with the discussion of case (i), where 
the combined symmetry $\hat{G}=G\T=G(k)\K\i$ acts  on $\H (k)$ as
\beq [\hat{G}(k),\H(k)]=0. \label{Magnetic-nonsymm}\eeq
Inserting Eq.(\ref{Chiral-form}) into Eq.~\eqref{Magnetic-nonsymm}, one obtains
$e^{ik}q(k)=q(-k)$,
and 
$ q_{n}=q_{-(n+1)}$ by use of $q(k)=\sum_n q_n e^{ink} $,
which implies
\beq q(k)=\sum_{n=0}^{\infty} q_n(e^{ink}+e^{-i(n+1)k}). \label{T-q-relation}\eeq
We observe that due to the symmetry constraint Eq.~(\ref{Magnetic-nonsymm}),
$q(\pi)=0$, i.e., there is a band crossing point at $k= \pi$. Note that the 
term $f_z(k)\sigma_3$ is not forbidden by the combined symmetry $G \T$.
However, Eq.~\eqref{Magnetic-nonsymm} requires that $f_z(k)$ is an odd function of $k$. Hence  $f_z(k)\sigma_3$ must vanish
at the high-symmetry points $k=0$ and $k=\pi$, due to the periodicity of the BZ, i.e., $f_z(k)=f_z(k+2\pi)$.
To summarize, the magnetic nonsymmorphic symmetry $G \T$ is sufficient to enforce the existence of a band crossing at $k=\pi$.

Next, we discuss choice (ii) for $\T$, in which case the combined symmetry $\hat{G}' \equiv G\T = G\sigma_3\K\i$ 
acts on $\H(k)$ as
\beq [\hat{G}'(k),\H(k)]=0 \label{action_onH_caseII}. \eeq 
Combining Eq.~\eqref{Chiral-form} with Eq.~\eqref{action_onH_caseII} we find that
$e^{ik}q(k)=-q(-k)$ .
Hence, the Fourier components $q_n$ must satisfy $q_n=-q_{-(n+1)}$, which yields
\beq q(k)=\sum_{n=0}^{\infty} q_n (e^{ink}-e^{-i(n+1)k}) . \label{q_expr_case_II} \eeq
Since $q(0)=0$ in Eq.~\eqref{q_expr_case_II}, there is an unavoidable band 
crossing at $k=0$.
As before, the term $f_z (k) \sigma_3$ is symmetry allowed, with $f_z(k)$ an odd function. Hence,
$f_z (k) \sigma_3$ vanishes at the high-symmetry points $k=0$ and $k=\pi$, and therefore cannot gap out
the band crossing point.


\begin{table}
	\begin{tabular}{ c| c |c }
		 Position & $G$,$\P$ & $G\T$ \\
		\hline
		$k=0$ & $\P G=-G^{\dag}\P$ & $\{G,\T\}=0$ \\
		$k=\pi$ & $\P G=G^{\dag}\P$ & $[G,\T]=0$ \\
	\end{tabular}
	\caption{
	The positions of the band crossings in the BZ are determined by the algebra of the symmetry operators. 
	 \label{Positions}}
\end{table}

From the above discussion, one infers that the commutation relation between
 $\T$ and $G(k)$ determines the position of the band-degeneracy points.
Namely, for $[G,\T]=0$ [case (i)] and $\{G,\T\}=0$ [case (ii)], we have  $(G\T)^2=+ e^{-ik}\sigma_0$
and  $(G\T)^2=- e^{-ik}\sigma_0$, respectively. Hence, 
we find that for case (i) $(G\hat{T})^2=- 1$ at $k=\pi$, while for case (ii) 
$(G\hat{T})^2=- 1$ at $k=0$.
Since $G\T$ is an anti-unitary operator, $(G\hat{T})^2=-1$ leads to a band degeneracy, in
analogy to  Kramers theorem.
Thus, for $[G,\T]=0$  the band-crossing point is at $k=\pi$, while for $\{G,\T\}=0$ it is at $k=0$, see Table~\ref{Positions}.

We note that a number of recent works~\cite{nonsymmorphic_Sato,New_crystalline_Fang,shiozaki_gomi_arXiv_16,Bernevig-Hourglass} have discussed edge band structures of two-dimensional (2D) nonsymmorphic insulators that are similar to the 1D bulk band structures studied here. 
In contrast to conventional topological insulators,  the edge bands of these 2D nonsymmorphic insulators do not connect valence and conduction bands. Hence, our results for the bulk band structures of 1D systems, can be applied directly to the edge spectrum of these 2D nonsymmorphic insulators. This suggests, in particular, that the crossing of the edge bands of these 2D systems is, at least in some cases, enforced by the nonsymmorphic symmetry of the edge theory.

In closing this section, we note that the existence
of a band crossing cannot be enforced by a nonsymmorphic
particle-hole symmetry, which is discussed in detail in Appendix~\ref{Appendix}.

%
%


\section{Topological classification of band-crossing points}
Let us now derive the  classification of the global  
topological properties of the considered Hamiltonians. By global topology, we mean a band structure is allowed to be deformed smoothly in the whole momentum space with the symmetries being preserved, in contrast to the ordinary local one, where deformations are restricted only in a open neighborhood around the band-crossing point. The group structure is given by the direct sum of the Hamiltonians.
To that end, we study whether the band crossings of the doubled Hamiltonians $\H\otimes\tau_0$ and $\H\otimes\tau_3$
can be gapped out by symmetry-preserving terms.
Here, $\tau_\mu$'s represent an additional set of Pauli matrices and $\tau_0$ is the $2 \times 2$ identity matrix. 
The symmetry operators for the doubled Hamiltonians are $G(k)\otimes\tau_0$, $\P\otimes\tau_0$, and $\T\otimes\tau_0$.
We observe that $\textrm{diag} (\H, \lambda\H)$ can be continuously deformed to $\textrm{diag} (\H, -\lambda\H)$  without breaking the symmetries and
without opening a gap at $k=0$ or $\pi$. In addition, we find that $m\sigma_0\otimes\tau_1$ is a symmetry preserving
mass term that gaps out the spectrum of $\H\otimes\tau_3$ in the entire BZ. 
It follows that the global topological features of $\H$ possess a $\Z_2$ classification, namely even number of copies can be gapped with symmetries being preserved, while odd number cannot. It is noted that although in the above simple deformation the gap is fully closed at $\lambda=0$, a more carefully chosen deformation can be made such that at any intermediate stage band crossing happens only at finite number of momenta.



To infer the classification of the local topological properties of the band crossing points, we enclose the degeneracy point by 
an  $S^0$ sphere (consisting of two points on the left and right of the degeneracy point)
and consider adiabatic deformations that do not close the gap on the chosen $S^0$.
The only possible gap opening term is $f_z (k) \sigma_3$, which however vanishes 
at the high-symmetry points $k=0,\pi$ due to the nonsymmorphic symmetry. This also
holds for multiple copies of $\H$. From this we conclude that the local topological features
of the band crossing points exhibit a $\Z$ classification.



\section{Higher-dimensional generalizations}
Our  results for symmetry-enforced band crossings in 1D
can be readily generalized to higher dimensions. 
We assume that the fractional translation is along the $k_x$ direction for $d$-dimensional systems.
The $d$-dimensional Hamiltonian $\H(k)$ can then be decomposed into a family of 1D Hamiltonians $h_{k_{\perp}}(k_x)=\H(k_x,k_{\perp})$, which are parametrized by the ($d-1$) momenta $k_{\perp}$ perpendicular to $k_x$. 
Let us briefly discuss how the three different types of nonsymmorphic symmetries that we considered above constrain this $d$-dimensional Hamiltonian.
(i) If $\mathcal{H}(k)$ is invariant under a unitary nonsymmorphic symmetry $G(k_x)$, then there are in general several branches of Fermi surfaces
(possibly of dimension $d>0$) that are parametrized by $k_{\perp}$.
(ii) If there exists in addition a reflection symmetry reversing $k_x$, 
then the Fermi surfaces are pinned at $k_x=0$ or $k_x = \pi$, depending on the algebraic relations obeyed by the symmetry operators,
as specified in Table~\ref{Positions}.
(iii) If we consider symmetries that relate $k$ to $-k$, such as a nonsymmorphic magnetic symmetry (or an additional inversion), 
then there exist 
$2^{d-1}$ 1D inversion invariant subsystems $h_{k_{\perp}^a}(k_x)$ ($a=1,\cdots, 2^{d-1}$) of $h_{k_{\perp}}(k_x)$ , 
which are labeled by the perpendicular momenta $k_{\perp}^{a}$ that are invariant under $k_{\perp}\rightarrow -k_{\perp}$. 
These subsystems have band degeneracies at $k_x=0$ or $k_x=\pi$, as determined by the algebraic relations in Table~\ref{Positions}.
The other 1D subsystems $h_{\tilde{k}_\perp}$,  where $\tilde{k}_\perp$ is \emph{not} invariant under $k_{\perp}\rightarrow -k_{\perp}$,
are generally gapped.


\subsection{Examples}
\begin{figure}
	\includegraphics[scale=0.8]{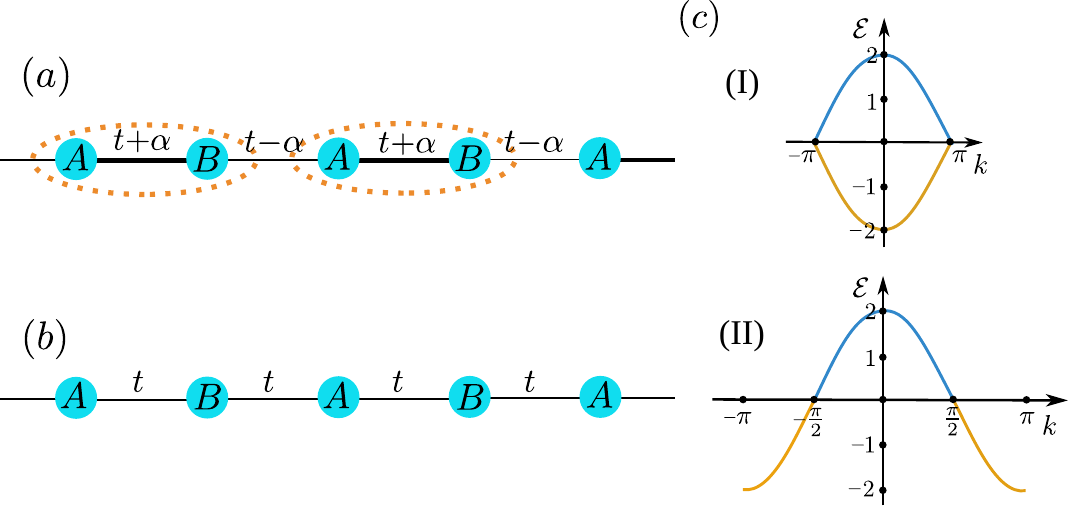}
	\caption{Illustration of the SSH model. (a) For $\alpha \ne 0$ the SSH model breaks the nonsymmorphic symmetry $G(k)$,
	Eq.~\eqref{Nonsymm-operator}. (b) At the critical point $\alpha =0$ the translation symmetry is promoted from $2\mathbb{Z}$ to  $\mathbb{Z}$, such that the nonsymmorphic symmetry G(k) is satisfied. 
	(c) Energy spectrum at $\alpha=0$ in  (I) the original BZ and (II) the unfolded BZ. \label{SSH}}
\end{figure}

We illustrate our theoretical results by two  examples: (i) The Su-Schrieffer-Heeger (SSH) model~\cite{heegerRMP88}
and (ii) the $\pi$-flux square lattice model.
The Hamiltonian of the SSH model is given by
\begin{equation}
\mathcal{H}_{\textrm{SSH}}(k)=\begin{pmatrix}
0 & \Delta(k)\\
\Delta^\dagger(k) & 0
\end{pmatrix},
\end{equation}
where $\Delta(k)=(t+\alpha)+(t-\alpha)e^{-ik}$, see Fig.~\ref{SSH}(a).
At the critical point $\alpha=0$ the system is invariant under a half translation followed by an exchange of $A$ and $B$ sublattices, which 
corresponds to the nonsymmorphic symmetry $G(k)$, Eq.~\eqref{Nonsymm-operator}, see Fig.~\ref{SSH}.
Since the SSH model also has the inversion symmetry $\P=\sigma_1\i$, we find in accordance with Table~\ref{Positions}
that there is a gapless point at $k= \pi$. Observe that
at $\alpha=0$ the translation symmetry is promoted from $2\mathbb{Z}$ (with translator $2a$) to $\mathbb{Z}$ (with the translator $a$) and
the SSH model becomes a 1D tight-binding model of free fermions with the dispersion  $\mathcal{E}(k)=2t\cos(k)$ [see Fig.~\ref{SSH}(c)]. It is noted that in order to create a bandcrossing point protected by nonsymmorphic symmetry in a tight-binding model through dimerization, one must ensure that the original non-dimerized model has \emph{two} chiral gapless modes. A nonvanishing $\alpha$, on the other hand, reduces the translation symmetry from $\mathbb{Z}$ to $2\mathbb{Z}$ and breaks the nonsymmorphic symmetry $G(k)$, leading to a topological ($\alpha < 0$) or trivial ($\alpha>0$) insulator, depending on the sign of $\alpha$.

\begin{figure}[h]
	\includegraphics[scale=0.48]{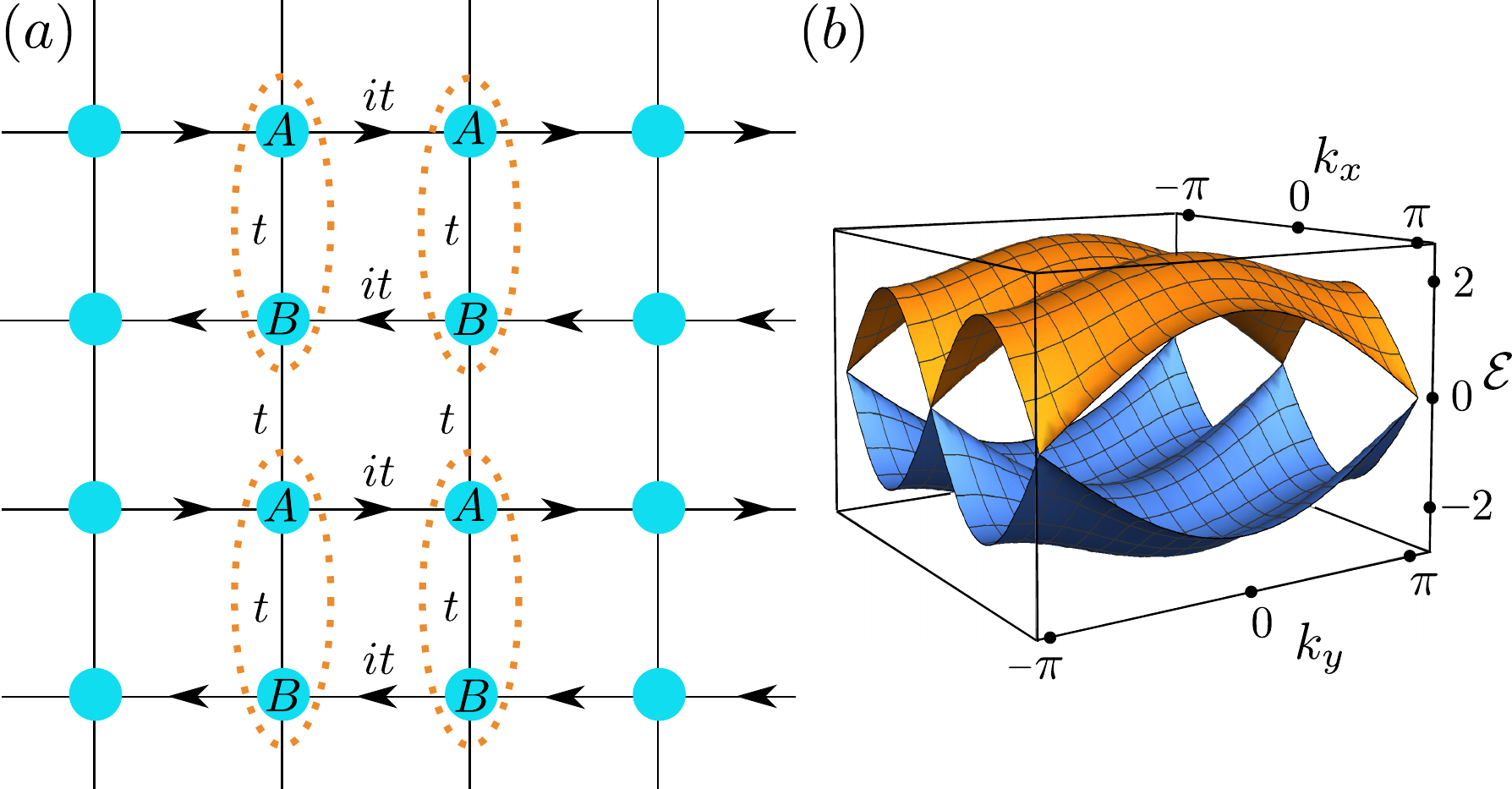}
	\caption{(a) Illustration of the $\pi$-flux square lattice model. This model is invariant under the 
		nonsymmorphic magnetic symmetry $G(k_y)\hat{\mathcal{K}}\hat{i}$, namely a time-reversal symmetry followed by the nonsymmorphic symmetry $G(k_y)$. (b) Energy spectrum of the $\pi$-flux square lattice model.
		\label{Half-flux}}
\end{figure}

The Hamiltonian of the $\pi$-flux square lattice model reads in momentum space
\begin{equation}
\mathcal{H}(k)=\begin{pmatrix}
2t\sin k_x & t+te^{-ik_y}\\
t+t e^{iky} & -2t \sin k_x
\end{pmatrix} ,
\end{equation}
where $t$ denotes the nearest neighbor hopping amplitude, see Fig.~\ref{Half-flux}(a).
The model is invariant under the nonsymmorphic magnetic symmetry $G(k_y)\hat{\mathcal{K}}\hat{i}$, 
which corresponds to a time-reversal symmetry $\T=\K\i$ followed by a half translation along $y$ and an exchange of $A$ and $B$ sublattices.  
The high-symmetry 1D subsystems $k_x=0$ and $k_x=\pi$  have gapless Dirac points at $k_y=\pi$, as shown in Fig.~\ref{Half-flux}(b). 
This is in agreement with Table~\ref{Positions}, since  $[G(k_y),\hat{\mathcal{K}}\hat{i}]=0$. Similar to the SSH model, the $\pi$-flux  state
is driven into a topological or trivial insulating phase by a dimerization $\alpha$ along $y$, that breaks the nonsymmorphic symmetry $G(k_y)$.
In closing, we observe that the $\pi$-flux model can be viewed as a higher-dimensional generalization of the SSH model.

\section{Discussions about multi-band theories \label{Discussion}}
It is noted that the topological arguments in this work are not limited to two-band models, which is exemplified by two cases in follows. First, if a multi-band theory has a chiral symmetry, then the Hamiltonian can be anti-diagonalized with the upper-right entry being a matrix $\Delta(k)$, and equation \eqref{Non-sym-q} still holds for $q(k)=\mathrm{Det}(\Delta(k))$. The topological argument around Eq.\eqref{zFz} implies that $\mathrm{Det}(\Delta(k))$ has to vanish somewhere in momentum space, namely there exists band-crossing points enforced by the chiral and nonsymmorphic symmetry.

Secondly, let us extend the spinless time-reversal symmetry discussed in Sec.\ref{MagneticNonsymmorphic} to the spinful one $\T=-i\sigma_2\K\i$, which acts in spinful four-band theories. If both $\T$ and $G$ are preserved ($G$ acts on the space of $\tau$), equations \eqref{Non-sym-q} and \eqref{T-q-relation} hold for $q(k)=\mathrm{Det}(\Delta(k))$, which implies $\mathrm{Det}(\Delta(k))$  vanishes at $k=\pi$. But diagonal terms have to vanish at $k=0$ and $\pi$ as required by the symmetries, except the chemical potential term.  Thus bands are enforced to cross at $k=\pi$. Note that the vanishing of $\mathrm{Det}(\Delta(k))$ at $k=\pi$ needs only the combined symmetry $G\T$, but both $\T$ and $G$ are required for that of the diagonal terms.

\begin{acknowledgments}
The authors thank C.-K.~Chiu for useful discussions. 
\end{acknowledgments}

	
\bibliography{TOPO3_v14}

\hfill

\appendix

\section{Nonsymmorphic Particle-hole symmetry}\label{Appendix}

In this section, we discuss whether the existence of a band crossing can also be enforced by a nonsymmorphic particle-hole symmetry.
As it turns out this is not possible. To see this, we consider for instance
an anti-unitary non-symmorphic particle-hole symmetry $\hat{G}''=G(k)\K$, 
which acts on $\H$ as
\begin{eqnarray}
G(k)\H(k) G^{-1}(k)=-\H^*(-k). 
\end{eqnarray}
Hence, the off-diagonal component of $\H (k)$ must satisfy
$ e^{ik}q(k)=-q(-k)$, and its Fourier components obey $q_n=-q_{-(n+1)}$.
Therefore, $q(k)$ can be written as
\beq q(k)=\sum_{n=0}^{\infty} q_n(e^{ink}-e^{-i(n+1)k}), \eeq
which vanishes at $k=0$. However, there does not exist a symmetry protected band crossing at $k=0$, since  the gap opening term
$z(k) \sigma_3$, with  $z(k)=z(-k)$ an even function, preserves the non-symmorphic particle-hole symmetry $\hat{G}''$.
To protect the band crossing point at $k=0$, an additional symmetry is needed which forbids the $z(k) \sigma_3$ term. 
For example, the chiral symmetry $\{\H,\S\}=0$,  with $\S=\sigma_3$  and  $\{\S, \hat{G}'' \}=0$, prevents
the  $z(k) \sigma_3$ term.

\end{document}